\documentclass[aps,pre,twocolumn,showpacs,floatfix,preprintnumbers,amsmath,amssymb,superscriptaddress,reprint]{revtex4-1}
\usepackage{amsmath}
\usepackage{amssymb}
\usepackage{graphicx}
\usepackage{epstopdf}
\usepackage{epsfig}
\usepackage{dcolumn}
\usepackage{bm}

\begin{document}
\title{Non-monotonic diffusion rates in atom-optics L\'{e}vy kicked rotor}
\author{Sanku Paul} 
\affiliation{
	Max-Planck-Institut f\"ur Physik komplexer Systeme, N\"othnitzer Stra\ss e 38, 01187 Dresden, Germany.}
\author{Sumit Sarkar} 
\affiliation{
	Indian Institute of Science Education and Research, Homi Bhabha Road, Pune 411 021, India.}
\author{Chetan Vishwakarma} 
\affiliation{
	Indian Institute of Science Education and Research, Homi Bhabha Road, Pune 411 021, India.}
\author{Jay Mangaonkar}
\affiliation{
	Indian Institute of Science Education and Research, Homi Bhabha Road, Pune 411 021, India.}
\author{M. S. Santhanam}
\affiliation{
	Indian Institute of Science Education and Research, Homi Bhabha Road, Pune 411 021, India.}
\author{Umakant Rapol}
\affiliation{
	Indian Institute of Science Education and Research, Homi Bhabha Road, Pune 411 021, India.}

\date{\today}
\begin{abstract}
The dynamics of chaotic Hamiltonian systems such as the kicked rotor
continues to guide our understanding of transport and localization processes.
The localized states of the quantum kicked rotor decay due to decoherence effects
if subjected to stationary noise. The associated quantum diffusion increases monotonically
as a function of a parameter characterising the noise distribution. In this work,
for the Levy kicked atom-optics rotor, it is experimentally shown that by tuning
a parameter characterizing the Levy distribution, quantum diffusion displays
non-monotonic behaviour. The parameters for optimal diffusion rates are analytically
obtained and they reveal a good agreement with the cold atom experiments and numerics.
The non-monotonicity is shown to be a quantum effect that vanishes in the classical limit.

\end{abstract}

\maketitle

Chaotic Hamiltonian systems continue to open up novel scenarios for 
momentum and energy transport in both the classical and quantum regimes.
The kicked rotor system, a particle periodically kicked by an external sinusoidal
field, is a paradigm for Hamiltonian chaos.
This sets standard benchmark for momentum transport, namely that, in the regime
of sufficiently strong kicking strengths, the onset of quantum interference effects
strongly attenuates the classical diffusive transport \cite{stockmann, haake}. This is the dynamical localization
scenario in which the system settles to a quasi-steady state and does not absorb energy
anymore. In contrast to this, novel transport scenarios have been exemplified by several
variants of kicked rotor. These include atom-optics based experimental realizations and
theoretical studies of directed transport in parity broken \cite{hainaut}, PT symmetric \cite{longhi}
and dissipative \cite{ggcarlo} kicked rotors. Anomolous transport has been observed in 
Levy kicked \cite{sumit}, relativistic \cite{mat, gong1} and a non-smooth version \cite{sanku}
of kicked rotor, while suppression of quantum diffusive transport was observed in 
higher dimensional \cite{manai}, non-ideal \cite{revu}, coupled \cite{simone, qin} 
and relativistic \cite{rozenbaum} kicked rotors. As the quantum kicked rotor 
is related to the Anderson model \cite{fishman, ander} for charge transport in a crystalline
lattice, all these results have applications for a larger class of disordered conductors
and time-dependent problems in condensed matter physics.

Kicked rotor is suitable for studying decoherence or the quantum to classical
transition of its localized states, especially since the classical and quantum 
signatures of transport are markedly distinct. In the classical domain, the temporal
evolution of mean energy is $\langle E \rangle = Dt$ where 
the diffusion coefficient $D \approx K^2/2$ and $K$ is the kick strength \cite{stockmann}.
In the corresponding quantum regime  $\langle E \rangle$ becomes asymptotically 
time independent, i.e., $\langle E \rangle = D (1 - \exp(-t/t_*)$
where $t_*$ is the Ehrenfest time-scale over which quantum dynamics follows the
classical behaviour. Thus, the numerical values of the kick strength $K>>1$ and
kick period $T$ determine the classical and quantum diffusion rates. In particular,
varying $K$ does not alter the qualitative nature of diffusion except if the
accelerator modes are present in the classical phase space \cite{iomin, arcy}. On the other hand,
if the parameters $K$ and/or $T$ are subjected to stationary noise, i.e, $K$ is replaced
by $K + \delta K$, where $\delta K$ is drawn from a stationary probability distribution,
then both theory and experiments have shown that quantum localization is not sustained \cite{klap, steck}.
A similar scenario unfolds if $T$ is subject to an additive noise \cite{oskay}.
In general, the strength of noise serves as a tunable parameter obtained from
noise characteristics and increasing it leads to quantum diffusion approaching its classical
limit in a monotonic fashion.

In many situations in which conductivity, and not localization, is desired the ability to tune
for optimal transport with a fixed kicked strength is useful. From the point of view of
atom-optics experimental realizations of kicked rotor, increasing the kick strength requires
improved hardware such as additional laser power that may not be always feasible.
In this work, we propose a mechanism based on tuning a parameter associated with
Levy-noise characteristics superposed on the kick period $T$ to obtain optimal momentum 
transport in an atom-optics kicked rotor system. The optimal diffusion coefficient, as
function of a parameter characterising the noise distribution function, is
analytically obtained and it is demonstrated through atom-optics based experiment as well.

\begin{figure*}[t]
\includegraphics*[width=\textwidth]{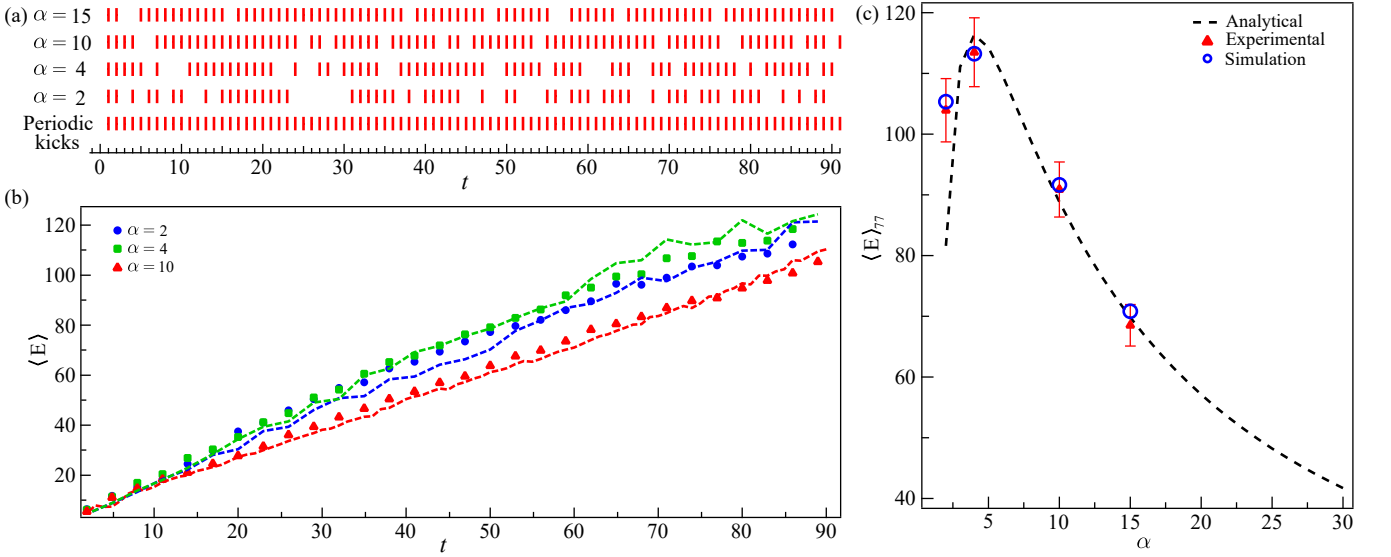}
\caption{(a) Schematic diagram representing the sequence of kicks for different $\alpha$.  (b) Quantum mean energy growth of the system for $\alpha=2$, $\alpha=4$ and $\alpha=10$ with  $K=6.0$, $\hbar_s=2.0$ . Symbols represent experimental data and solid lines are the numerical results. Experimental data has $\pm$5\% uncertainty (not shown in figure). (c) Represents the behavior of quantum diffusion coefficient by plotting energy after 77 kicks as a function of $\alpha$ for $K=6.0$, $\hbar_s=2.0$. Triangular symbols with error bars represent the experimental data, Circular symbols are the numerically calculated values and dashed line is the plot of  analytical expression given in Eq. (\ref{eq_10}) for the break-time $t^{*}\approx 5$, as achieved in our experiment.}
\label{alpha}
\end{figure*}

We consider the dimensionless Hamiltonian of the atom-optics quantum kicked
rotor system given by
\begin{equation}
\widehat{H} = \widehat{p}^{2}/2 + K \cos \widehat{x} \sum_n (1-g_n) \delta(t-n).
\label{eq_1}
\end{equation}
In this, $g_n$ is a stochastic variable that controls if an external field of kick strength $K$
is applied to the cold atomic cloud at $n$-th time instant. Further, $g_n$ is
taken from a discrete Bernoulli distribution such that if $g_n=0$, the particle 
experience a kick and if $g_n=1$ no kick is applied.
The waiting time between the occurrences of $0$ is drawn from a L\'{e}vy waiting time
distribution $w(\tau)\sim \tau^{-1-\alpha}$, where $\alpha$ is the L\'{e}vy exponent \cite{lutz, zas1}.
The regime of $0< \alpha < 1$ corresponds to diverging mean waiting time $\bar{\tau}$ and, as we had
demonstrated earlier, this effectively leads to slower decay of decoherence \cite{sumit}.
In this paper, we exploit the dynamics when the kicks are imparted at time intervals
governed by $w(\tau)$ with exponent $\alpha >1$ and $\bar{\tau}= \frac{\alpha}{\alpha-1}$ is well defined.
In this regime, we demonstrate through both theory and experimentation the existence of 
an optimal quantum diffusion as a function of $\alpha$.
This optimality is an unusual property since for other commonly used additive noise sources
such as white or Gaussian noise, quantum diffusion is known to be usually monotonic as a function 
of noise strength \cite{schom, lutz}.

The quantum dynamics of the system in Eq. (\ref{eq_1}) for $\alpha>1$ is studied
using the Floquet analysis. The Floquet operator can be written as
        \begin{equation}
        \begin{aligned}
        F(K_n) &= e^{-ip^2/2\hbar_s} ~ e^{-iK \cos x/\hbar_s } ~ e^{-iK_n^{'} \cos x/\hbar_s },\\
         &= F(K)~F(K_n^{'}).
        \end{aligned}
        \label{eq_2}
        \end{equation}
where $\hbar_s$ is the scaled Planck's constant, $K_n=K (1-g_n)$ and $K_n^{'}=-K g_n$. 
Further, $F(K)$ represents the Floquet operator of the standard kicked rotor 
$F(K) = e^{-ip^2/2\hbar_s} ~ e^{-iK \cos x/\hbar_s }$ for which $g_n=0$ for all $n$. 
The noisy rotor corresponds to $F(K_n^{'})$.
The starting point of the analysis are the eigenstates $|r\rangle$ of the Floquet 
operator $F(K)$ given by
        \begin{equation*}
        F(K) | r \rangle = e^{-i \eta_r} | r \rangle,
        \label{eq_4}
        \end{equation*}
where $\eta_r$ is the quasi-energy of the state $|r\rangle$. For $K>>1$ as we have taken,
$\eta_r$ would be the localized states. The factor $g_n$ induces
noise in the kicking sequence whenever $g_n=1$ and the kicks are not imparted at those
time instants. Thus, under the action of $F(K_n^{'})$ system transitions from state $\eta_r$.
The survival probability amplitude $A_{r} (t^{'},t^{''})$ of the noisy system to remain
in the state $| r \rangle$ in a given time interval $[t^{''},t^{'})$ \cite{lutz} is
        \begin{equation}
        A_{r} (t^{'},t^{''}) = {\mathcal N} ~\left\langle r\left | \mathcal{T} \prod_{n=t^{''}}^{t^{'}-1}F(K_n)\right | r \right \rangle,
        \label{eq_5}
        \end{equation}
in which $\mathcal{T}$ represents time ordering and ${\mathcal N}=e^{-i \eta_r (t^{'}-t^{''})}$
is introduced to normalize the survival probability amplitude for the noiseless system.

        By using random-phase approximations and performing an average over all the quasi-energy
states and also over the random phases of the initial state, the survival probability amplitude
to remain in state $| r \rangle$ for $\alpha>1$ takes the form 
        \begin{equation}
        A_{r} (t^{'},t^{''}) = q \left( K_n^{'}/ \hbar_s \right)^{G(t^{'},t^{''})}.
        \label{eq_6}
        \end{equation}
In this, $G(t^{'},t^{''})$ represents the number of noisy events in the interval $[t^{''},t^{'})$ and
        \begin{equation}
        \begin{aligned}
        q \left( K_n^{'}/ \hbar_s \right) = & 1 - (K_n^{'2}/2!~\hbar_{s}^{2})~\overline{ \cos^2(x) }  +\\
         &(K_n^{'4}/4!~\hbar_{s}^{4}) ~\overline{ \cos^4(x) }+\cdots.
         \end{aligned}
       \label{eq_7}
        \end{equation}
Thus, it can be inferred that $\left|q \left( K_n^{'}/ \hbar_s \right)\right| <1$, indicating
that the survival probability in the state $|r\rangle$ decays over time and the consequent
state transitions result in diffusion. By using the force-force correlator which is related
to the decoherence factor \cite{lutz, cohen, schriefl}, the mean energy for $\alpha>1$ can be obtained as
\begin{equation}
\begin{aligned}[b]
\langle E\rangle_t &\sim \frac{K^2}{2}\left(1-e^{\frac{-t}{t^{*}}}\right)  -\frac{K^2}{2\alpha}t
   + \frac{K^2}{2} \frac{t}{1-\frac{\alpha}{c}} + O\left(a^{t}\right)\\
   &\sim I_1 + I_2 + I_3 + O\left(q^{t}\right).
\end{aligned}
\label{eq_8}
\end{equation}
where $c=(q^2-1) t^{*}$ and $t^{*}\sim \frac{K_n^{2}} {\hbar_s^2}$ is the break-time 
of the standard kicked rotor. To physically understand this expression, we analyse each of these
terms. The first term $\left( I_1 \left(=\frac{K^2}{2}\left(1-e^{\frac{-t}{t^{*}}}\right)\right) \right)$
represents the energy growth of the standard kicked rotor, $I_2$ $\left(=\frac{K^2}{2\alpha}t\right)$ 
corresponds to the missing of kicks and the energy growth is represented by
$I_3 + O\left(q^{t}\right)$ $\left(= \frac{K^2}{2} \frac{t}{1-\frac{\alpha}{c}} + O\left(a^{t}\right)\right)$
results from decoherence due to the introduction of noise.

\begin{figure}[t]
\includegraphics[width=0.49 \textwidth]{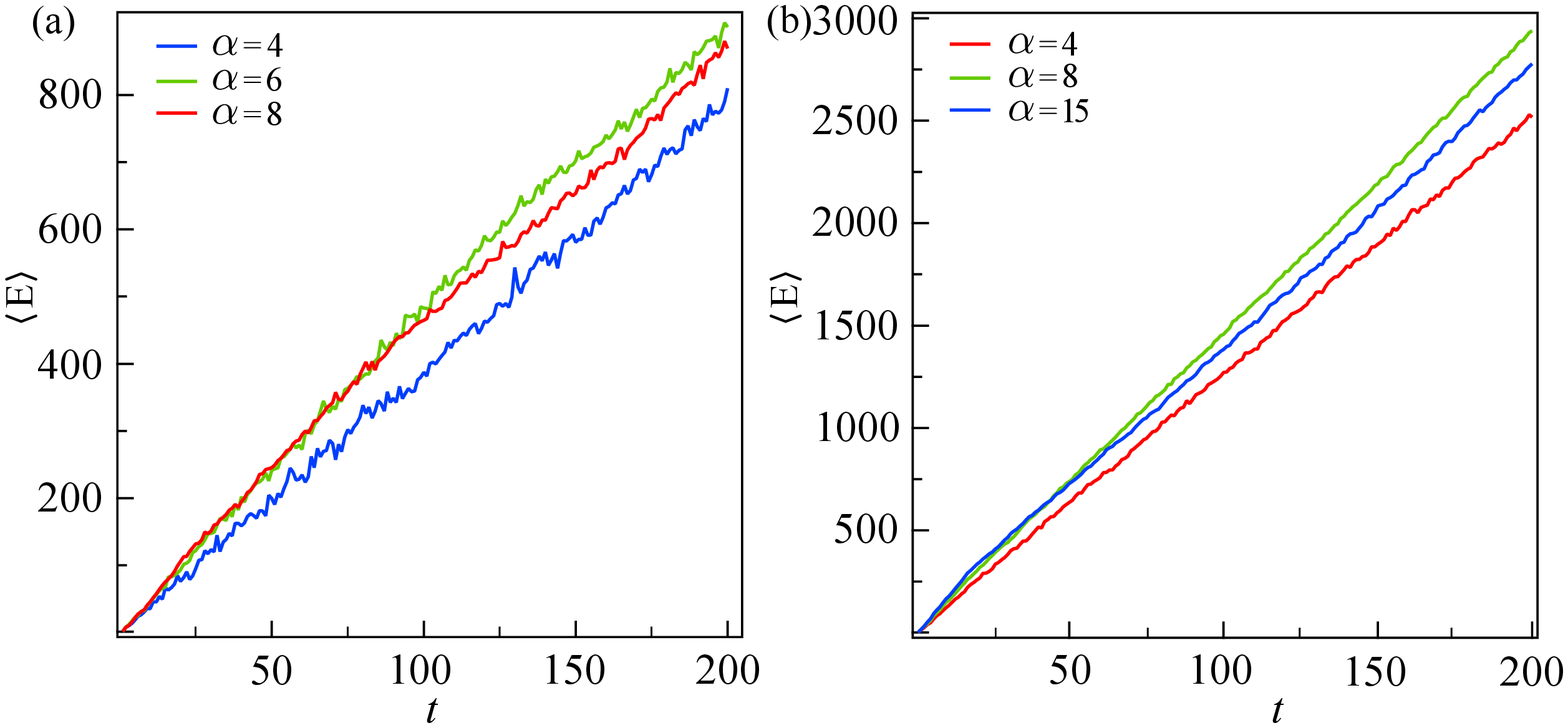}
\caption{Numerically calculated mean energy growth of the system for different $\alpha$ at 
(a) kick strength $K=10$, $\hbar_s=2$ and (b) kick strength $K=15$, $\hbar_s=2$.}
\label{ene2}
\end{figure}

After long times, $t \gg t^{*}$, Eq. (\ref{eq_8}) reduces to $ \langle E\rangle \sim D t$,
in which the diffusion coefficient $D$ is given by
        \begin{equation}
        \begin{aligned}[b]
        D ~ \sim& ~ \frac{K^2}{2}\left(-\frac{1}{\alpha}
         + \frac{1}{1-\frac{\alpha}{c}}\right). \\
        \end{aligned}
        \label{eq_10}
        \end{equation}
This reveals a nonlinear dependence of the diffusion coefficient $D$ on
both the kick strength $K$ and L\'{e}vy exponent $\alpha$.
The mean energy of the system grows linear in time
indicating a dominance of diffusion. In contrast, the diffusion coefficient $D$ has a quadratic
dependence on kick strength but significantly the dependence on $\alpha$ is not monotonic.
The expressions in Eqs. \ref{eq_8}-\ref{eq_10} form the central analytical results
of this paper. In what follows, we describe an atom-optics based experiment
to verify these results.

The experimental sequence is similar to that given in Ref. \cite{sumit}. We prepare a laser-cooled cloud of  $^{87}$Rb atoms in magneto-optical trap (MOT). This is followed by further forced evaporative cooling in a crossed optical dipole trap ($\lambda = 1064$nm).  The cold atomic ensemble has $\sim 2 \times 10^5$ atoms at temperature $\sim$ 3 $\mu$K and follows Maxwell-Boltzmann distribution in momentum space. This atomic ensemble serves as the initial Gaussian wavepacket for simulating the quantum kicked rotor. The process of kicking is implemented using a pulsating 1-D optical lattice.  The lattice laser beam is $\sim$-6.7 GHz detuned from the $|F = 1\rangle \longrightarrow |F^\prime = 2\rangle$ transition of $^{87}$Rb. The lattice beam is derived from the 1st order diffraction of an acousto-optic modulator (AOM). The lattice is turned ON and OFF by switching the RF power that drives the AOM via a high frequency switch. The pulse ON time for the applied kicks is $\approx$ 220 ns and the free propagation time is kept to be $\approx$10.6 $\mu$s. For the parameters used in the experiment, the scaled Planck’s constant is $\sim$ 2 and the kick strength(K) is calculated to be 6 with 10 \% uncertainty. For realization of the  L\'{e}vy noise, a sequence of wait times following  L\'{e}vy statistics are fed to an arbitrary waveform generator which in turn controls the RF switch of the AOM driver.  The presented experimental data for each  L\'{e}vy exponent is an average over five different noise realizations.

Figure \ref{alpha}(a) shows one realization of the actual kicking sequence used in
the experiment for several values of L\'{e}vy exponent $\alpha$. 
As $\alpha$ increases, the mean number of missed 
kicks decreases such that in the limit of $\alpha \to \infty$ it is effectively a periodic
kick sequence. In this limit, we expect the system to display the properties of
standard kicked rotor system, i.e, Eq. \ref{eq_1} with $g_n=0$ for all $n$.
In Fig. \ref{alpha}(b), the mean energy growth of the system for $K=6$,
$\hbar_s=2$ while the Levy exponent $\alpha$ is varied. In this, symbols with error bar
represent experimental data while dashed lines are the numerical results. It
can be immediately recognized that the diffusion shows an unusual property, namely,
non-monotonicity as a function of $\alpha$.
As $\alpha$ is tuned from 2.0 to 10.0, diffusion reaches a maximum at about $\alpha \approx 4.0$
and then begins to decrease.

To obtain a broader perspective of this result, for an arbitrarily chosen time $t=\bar{t}$,
the mean energy $\langle E \rangle_{\bar{t}}$ is tracked as a function of $\alpha$.
In Fig \ref{alpha}(b), $\langle E \rangle_{\bar{t}}$ at a fixed time of $t=77$  (in units
of kick period) is displayed as a function of $\alpha$
for the case of kick strength $K=6.0$ and $\hbar_s=2.0$. It is apparent that $D$ has
non-monotonic dependence on $\alpha$ such that its maxima occurs at $\alpha=\alpha_c$.
The symbols with error bars represent experimental data, and the dashed line is
the analytical expression in Eq. (\ref{eq_10}), in which the break-time, $t^{*}$ is treated
as a fitting parameter. The best estimate of the break-time $t^{*}=7.27$ (in units of kick period)
obtained through fitting is very close to the theoretical prediction 
$t^{*}\sim\frac{K^2}{\hbar_{s}^2}=9$.  Clearly, the experimental data displays an excellent 
agreement with the theoretical result in Eq. (\ref{eq_10}).
Fig. \ref{alpha}(c) displays an unusual feature that $\langle E \rangle_t$ , at $t=77$,
initially increases with $\alpha$ until $\alpha < \alpha_c$, and begins to decay 
for $\alpha > \alpha_c$.  In this case, the mean energy growth is maximum at 
$\alpha = \alpha_c \approx 4.0$.
The choice of $t=77$ is for illustrative purposes and a similar behaviour with identical
value of $\alpha_c$ is obtained for any other $t$, provided other parameters are held constant.

This can be understood as follows. Note that the contribution to mean energy due to
$I_2$ is negative and hence acts to suppress the growth of mean energy. 
On the other hand, $I_3 > 0$ and tends to increase mean energy. The competition between
these two terms leads to a maxima at $\alpha=\alpha_c$, as seen in Fig. \ref{alpha}(c).
Physically, $I_2$ originates due to the noisy kick sequence and is related to the 
probability of missing a kick at a given time instant. The exponent $\alpha$ present 
in $I_2$ controls the mean number of missed kicks. As $\alpha$ increases, fewer kicks
are missed, and hence the mean energy increases.
The term $I_3$ arises from decoherence of the localized state and is therefore associated 
with delocalization and consequent increase of mean energy. Since the signs of $I_2$ and 
$I_3$ are opposites of one another, the net
effect of these two competing terms leads to a maxima in mean energy growth
at $\alpha=\alpha_c$. Presence of a single maxima is due to monotonic behaviour of $I_2$ and $I_3$
respectively as a function of $\alpha$.

\begin{figure}[t]
\includegraphics[width=0.35\textwidth]{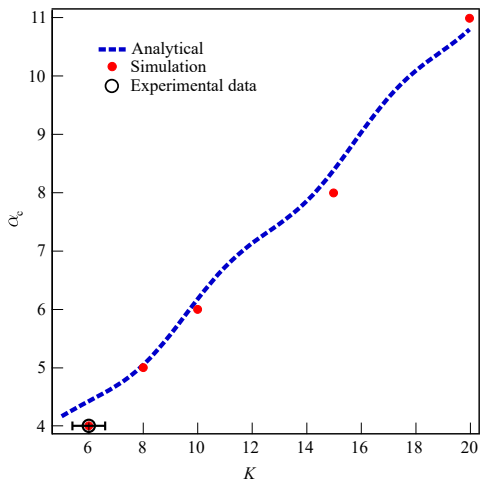}
\caption{The variation of critical exponent $\alpha_c$ as a function of
$\frac{K}{\hbar_s}$, where $\hbar_s=2$ and $K$ is a free variable. Dashed line is
the plot of analytical expression given in Eq.\ref{eq_9} and symbols represent
the values of $\alpha_c$ as observed in the numerical simulations. At $K=6$,
experimental data for $\alpha_c$ from Fig. \ref{alpha}(c) is also shown.}
\label{alphac}
\end{figure}

Further, we explore the limit of $\alpha \to \infty$. For any fixed value of $t$ such that
$t \gg t^{*}$ and now if the limit $\alpha \gg \alpha_c$ is taken, both $I_2$ and $I_3$ tends to zero.
This leads to $ \langle E \rangle \sim K^2/2$, a time-independent value that corresponding
to that of the localized state obtained with periodic kicking. For large $\alpha$,
the mean number of missed kicks becomes vanishingly small and hence the system essentially
works like the standard kicked rotor. As observed in Fig. \ref{alpha}(c), for large $\alpha$,
the mean energy is seen to be approaching a constant value.  This constant value of 
$K^2/2 = 18$ is approached very slowly since $D \propto \alpha^{-1}$ as $\alpha \to \infty$.


In Fig. \ref{ene2}(a,b), numerically simulated energy diffusion of the system in 
Eq. (\ref{eq_1}) is displayed for kick strengths $K=10$ and $K=15$ respectively. 
It can be noticed from this figure that $\alpha_c$ depends on the kick strengths. 
For $K=10$, $\alpha_c \approx 6$ and for $K=15$, we obtain $\alpha_c \approx 8$. 
These numerical estimates of $\alpha_c$ are in accordance with that predicted by
the analytical expression in Eq. (\ref{eq_9}).
Hence, the results shown in Fig. \ref{alpha}(b,c) repeates itself qualitatively for other 
values of kick strengths as well except that $\alpha_c$ at which the diffusion is
maximum depends on $K$. For a fixed value time $t=\bar{t}$, starting from Eq. \ref{eq_10}
an expression for $\alpha_c$ is derived by extremizing $D$ with respect to $\alpha$ and
it gives,
\begin{equation}
\alpha_c=\frac{1+\sqrt{(1-q^2) t^{*}}}{1-\frac{1}{(1-q^2) t^{*}}}.
\label{eq_9}
\end{equation}
In this, $q \equiv q(K_n^{'}/ \hbar_s)$ and $t^{*} \equiv t^{*}(K/ \hbar_s)$.
Hence, $\alpha_c$ depends only on the ratio $\frac{K}{\hbar_s}$. Figure \ref{alphac},
shows $\alpha_c$ as a function of $K$ for a fixed value of $\hbar_s$. In this figure, the 
result of Eq. \ref{eq_9} is matched against the numerical simulations of quantum 
Levy kicked rotor. The experimentally obtained data point for $K=6$ is also shown.
To a first approximation, $\alpha_c$ increases linearly with $K$ and the agreement with
simulation and experimental result is good. This result also emphasises the quantum nature
of the non-monotonic diffusion in Levy kicked rotor. As $\hbar \to 0$, in the semiclassical
limit, $K/\hbar >> 1$ and hence $\alpha_c \to \infty$. Hence, the non-monotonic diffusion
is a quantum phenomenon and cannot be seen in the classical Levy kicked rotor. Indeed,
in the classical numerical simulations (not shown here) of this system, the diffusion
is indeed monotonic for all values of $K$.

\begin{figure}[t]
\includegraphics[width=0.35\textwidth]{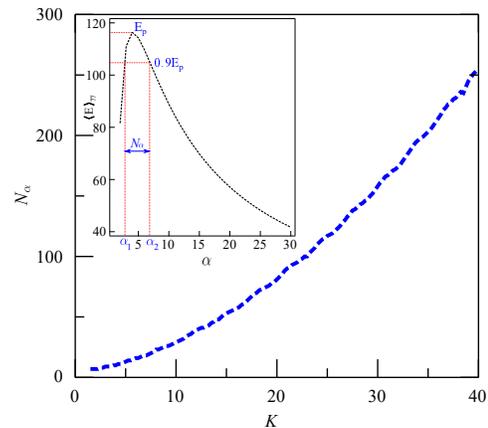}
\caption{Variation of  $N_{\alpha} (= \alpha_2 - \alpha_1)$ with kick 
strength $K$. $\alpha_1, \alpha_2$ are the values of $\alpha$ where the 
energy is 90\% of the energy at $\alpha_c$. Inset shows a replica of 
fig. 1(c) with $E_p$ being the energy at $\alpha_c$ and $N_{\alpha}$ shows 
the width in $\alpha$ at energy equals to 0.9$E_p$. }
\label{nalpha}
\end{figure}


Another feature that can be infered from Fig. \ref{ene2} is that for large
kick strengths such as $K \gtrapprox 10$, the curves for $\langle E \rangle_t$ 
corresponding to different values of $\alpha$ are tend to be close each other or
even overlap. However, for $K=6$ in Fig. \ref{alpha}, 
as $\alpha$ is varied $\langle E \rangle_t$ remains quite distinct. The extent to which the system
responds to variation in Levy exponent $\alpha$ can be quantified by a
"response" curve defined as follows. We define a "bandwidth" in $alpha$
space defined as $N_\alpha = \alpha_2 - \alpha_1$, in which $\alpha_1$ and
$\alpha_2$ are such that
\begin{equation}
\langle E \rangle_t(\alpha_1) = \langle E \rangle_t(\alpha_1) = 0.9 \langle E \rangle_t(\alpha_c).
\end{equation}
Thus, in analogy with $Q$-values of oscillators, smaller values of $N_\alpha$ would 
correspond to higher sensitivity of the system to changes in $\alpha$ in stark
contrast to  larger $N_\alpha$ corresponding to lower sensitivity.
Figure \ref{nalpha} shows $N_\alpha$ obtained through numerical simulations starting
from Eq. \ref{eq_8}.
The inset in this figure pictorially illustrates the definition of $N_\alpha$ for the data
shown in Fig. \ref{alpha}(c). It is seen that as $K$ increases, $N_\alpha$ increases
pointing to increasing loss of sensitivity to changes in $\alpha$ for large kick strengths.
This behaviour of $N_\alpha$ explains why $\langle E \rangle_t$ curves nearly 
overlap for large $K$. If $ K>>1$, there is a wide band of $\alpha$ values for which
diffusion rates are nearly same as that at $\alpha_c$.
Physically, it is reasonable to expect that large kick strengths are classically
chaotic regimes, and the kicked rotor is less sensitive to variations in $\alpha$.

In summary, the dynamics of L\'{e}vy kicked rotor system is studied through
experiments and simulations in the regime of L\'{e}vy exponent $\alpha > 1$. 
In this, instead of periodic kicking of the standard kicked rotor, the
system misses kicks at time intervals governed by the L\'{e}vy waiting time
distribution,$w(\tau) \sim \tau^{-1-\alpha}$. For $\alpha>1$, the mean waiting time 
$\bar{\tau}=\frac{\alpha}{\alpha-1}$. The central result is the non-monotonic
behavior of difusion coefficient upon variation of $\alpha$. In general, for any
value of kick strngth $K$ such that the system is classically chaotic, the diffusion
rate as a function of $\alpha$ displays a single maximum at $\alpha=\alpha_c$.
It is also shown that $\alpha_c$ is linearly dependent on $K/\hbar_s$, showing that
the non-monotonicity of diffusion is a quantum effect that vanishes in the classical limit. 
Non-monotonicity of diffusion is a surprising feature that is generally not
seen in the standard quantum kicked rotor and its variants.

Sanku Paul and Sumit Sarkar contributed equally to this work. Sanku Paul performed
theoretical and computational work. Sumit Sarkar was part of the experimental team.

\end{document}